\documentclass[sigconf]{acmart}
\usepackage{algorithm}
\usepackage{algorithmic}
\usepackage{multirow}
\usepackage{booktabs}
\usepackage{graphicx}
\usepackage{makecell}
\renewcommand\footnotetextcopyrightpermission[1]{} 
\begin{document}

\title{Efficient and Effective Retrieval of Dense-Sparse Hybrid Vectors using Graph-based Approximate Nearest Neighbor Search}


\author{Haoyu Zhang}
\affiliation{%
  \institution{Tsinghua University}
  \country{China}}
\email{hy-zhang22@mails.tsinghua.edu.cn}

\author{Jun Liu}
\affiliation{%
  \institution{Shanghai Jiaotong University}
  \country{China}}
\email{liu-j23@sjtu.edu.cn}

\author{Zhenhua Zhu}
\affiliation{%
  \institution{Tsinghua University}
  \country{China}}
\email{zhuzhenhua@mail.tsinghua.edu.cn}

\author{Shulin Zeng}
\affiliation{%
  \institution{Tsinghua University}
  \country{China}}
\email{zengshulin@mail.tsinghua.edu.cn}

\author{Maojia Sheng}
\affiliation{%
  \institution{ByteDance}
  \country{China}}
\email{shengmaojia@bytedance.com}

\author{Tao Yang}
\affiliation{%
  \institution{ByteDance}
  \country{China}}
\email{yangtao.chen@bytedance.com}

\author{Guohao Dai}
\affiliation{%
  \institution{Shanghai Jiaotong University}
  \country{China}}
\email{daiguohao@sjtu.edu.cn}

\author{Yu Wang}
\affiliation{%
  \institution{Tsinghua University}
  \country{China}}
\email{yu-wang@mail.tsinghua.edu.cn}








\begin{abstract}
The approximate nearest neighbor search (ANNS) over embedded vector representations of texts is a common method in information retrieval. Lexical-based sparse vectors and semantic-based dense vectors are two important and complementary information representations. The fusion of them has been empirically demonstrated to enhance effectiveness and robustness. The prevailing approach for hybrid vector search is to cunduct the sparse and dense vector search separately, then merge the results. However, the two-route method suffers from high system complexity and poor scalability. Another promising approach is to build unified index for hybrid vectors. Nevertheless, the unified system faces two key challenges regarding accuracy and efficiency. (1) The significant difference in data distribution between dense and sparse vectors poses a challenge in combining them, leading to low accuracy. (2) The high dimensionality and sparsity characteristics of sparse vectors cause expensive distance computation, resulting in inefficient retrieval.

To address the above challenges, we propose a novel graph-based ANNS algorithm for dense-sparse hybrid vectors. (1) For effectiveness, we propose the distribution alignment method by pre-sampling dense and sparse vectors and analysing the statistic of distance distribution. It can improve the accuracy by 1\%$\sim$9\%. (2) For efficiency, we find that the sparse vectors information has little contribution at the beginning of hybrid vectors index construction and search. It inspires us to design an adaptive two-stage computation strategy, which initially computes dense distances only and later computes hybrid distances. Further, we prune the sparse vectors to speed up the calculation. Compared to naive implementation, we achieve $\sim2.1\times$ acceleration. Thorough exhaustive experiments on mainstream text retrieval datasets, our algorithm achieves 8.9$\times$$\sim$11.7$\times$ end-to-end throughput at equal accuracy compared with existing hybrid vector search algorithms.

\end{abstract}





\maketitle

\section{Introduction}

Information retrieval focuses on locating relevant information resources, such as documents or passages, in response to user queries. It finds extensive application in various domains, including web search, recommendation systems, and knowledge mining \cite{zhao2022dense}. To be able to find relevant information from a large amount of candidate texts, a common practice is to map texts into vectors, and use the distance between vectors to represent the similarity between texts \cite{wong1987modeling}. Therefore, the information retrieval task is transformed into a nearest neighbor search (NNS) task on vectors.

\begin{figure}[!tp] 
  \centering
  \includegraphics[width=\linewidth]{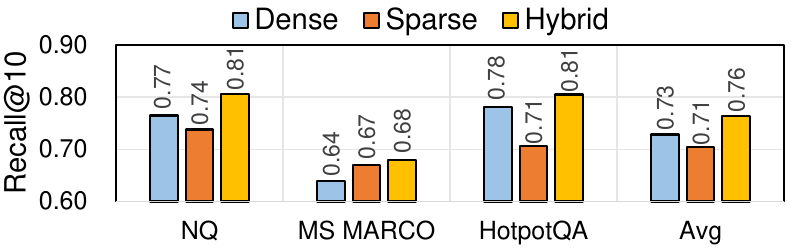}
  \caption{Comparison of recall accuracy (Recall@10) on NQ, MA MARCO, HotpotQA datasets, using dense (embedded by BGE model \cite{bge_embedding}), sparse (embedded by SPLADE model \cite{bge_embedding}), and hybrid vector retrieval approaches.}
  \label{fig: hybrid recall}
\end{figure}

Existing text embedding approaches can be categorized into sparse and dense methods.
Sparse methods, such as TF-IDF \cite{tf-idf} and BM25 \cite{bm25}, rely on keywords frequency statistics to map texts into high-dimensional sparse vectors. While these methods excel at exact keyword matching and offer interpretable representations, their performance suffers in scenarios with keyword mismatches.
On the other hand, dense methods, such as Word2Vec \cite{word2vec} and BERT-based DPR \cite{DPR}, learn low-dimensional dense vectors that encode semantic and contextual information. These vectors excel in semantic matching but lack the keyword information needed for precise matching.
To exploit the complementary strengths of both methods and mitigate their respective limitations, the dense-sparse hybrid method has been developed \cite{lee2023complementarity, gao2021complement, chen2022out, wang2021bert, luan2021sparse}. These hybrid methods have demonstrated significant improvements in retrieval accuracy, as shown in Fig. \ref{fig: hybrid recall}. Furthermore, they exhibit enhanced robustness in cross-domain and long-text retrieval scenarios \cite{chen2022out, luan2021sparse}. In the growing field of Retrieval Augmented Generation (RAG) of Large Language Models (LLMs), the adoption of hybrid retrieval techniques has emerged as a widespread practice.

Although dense-sparse hybrid vector retrieval has shown improved accuracy and robustness, there is a lack of established methods for a unified and efficient hybrid vector search. Efficiently searching in such hybrid spaces poses a challenge due to the lack of overlap between retrieval methods that excel with sparse or dense vectors individually. The classical sparse retrieval method, inverted indexing, faces issues with the dense vector part requiring brute-force search, leading to a 10$\sim$100$\times$ increase in latency. Conversely, for dense retrieval methods, the high-dimensional sparse vector poses challenges for index construction (e.g. linear increase of clustering time with dimension in Inverted File index (IVF) \cite{douze2024faiss}) and distance calculation (e.g. expensive sparse distance computation between points in graph-based index). 
To avoid the above challenges, the prevailing approach is a two-route retrieval method, \textit{i.e.}, conducting sparse and dense vector searches separately, then merging the results. However, this approach increases system complexity by requiring two indexes and data synchronization. It also constrains overall throughput and may compromise result accuracy as the top-k hybrid results may not be present in the individual sparse and dense top-k results.

\begin{figure}[!tp] 
  \centering
  \includegraphics[width=0.95\linewidth]{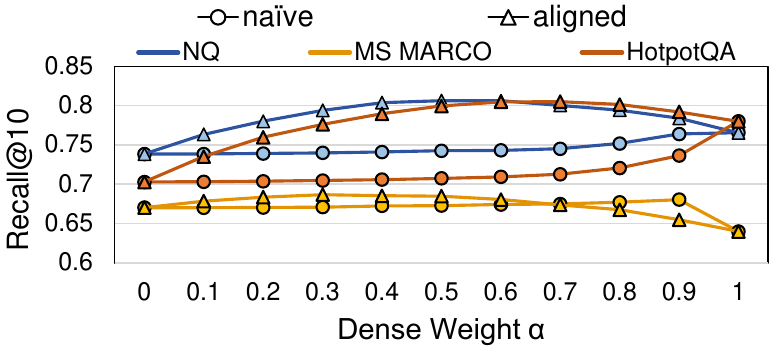}
  \caption{Recall accuracy (Recall@10) of hybrid search with varying dense weights on NQ, MS MARCO, and HotpotQA datasets, embedded by BGE and SPLADE models. After dense and sparse space alignment, the optimal weights shift to around 0.5, achieving higher recall accuracy.}
  \label{fig: dense weight}
\end{figure}

\begin{figure}[!tp] 
  \centering
  \includegraphics[width=\linewidth]{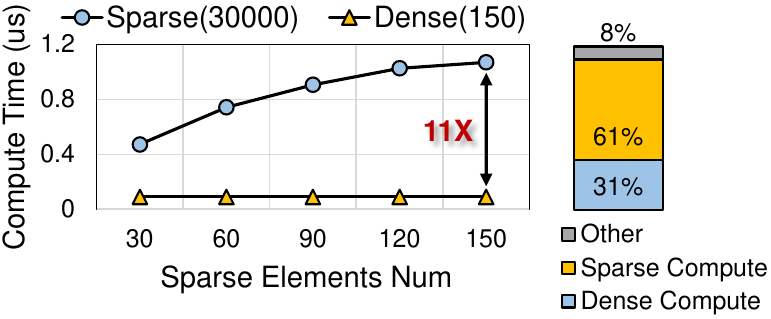}
  \caption{Left: Comparison of IP distance computing time of sparse vectors with varying number of non-zero elements (dim = 30,000) v.s. dense vectors (dim = 150). Right: Time distribution of hybrid search using naive HNSW algorithm. Sparse calculation accounts for $\sim$61\%.}
  \label{fig: sparse cost}
\end{figure}

Therefore, we require a unified dense-sparse hybrid vector search approach to ensure accurate results and efficient search performance. While Bruch et al. \cite{bruch2023bridging} have introduced a IVF-based hybrid vector retrieval algorithm, its efficiency remains suboptimal. Considering the high connectivity advantage of the graph-based index in high-dimensional spaces, we adopt the Hierarchical Navigable Small World (HNSW) \cite{malkov2018efficient}, one of the optimal graph-based ANNS algorithms, for hybrid vectors retrieval. However, applying HNSW directly to hybrid vectors created by concatenating sparse and dense vectors presents the following two challenges: 
\begin{enumerate}
\item The distance distributions of sparse vectors and dense vectors are inconsistent, which means disparate similarity differences represented by the same sparse and dense distance differences. This makes it difficult to optimally combine the two types of similarities and damaging retrieval \textbf{effectiveness}. We have observed that using linear weighted fusion of sparse and dense distances with varying weights has a substantial impact on accuracy, as shown in Fig. \ref{fig: dense weight}.
\item While graph index can maintain excellent performance in high-dimensional space, sparse vectors incur high computational overhead, degrading \textbf{efficiency} and even worse than dense recall at low accuracy. This is because sparse vectors are stored in a compressed indices-value format to save space, which introduces indices matching overhead during distance computations. We discovered that computing inner product (IP) distance between sparse vectors takes $\sim$11$\times$ longer time than dense vectors, and accounts for over 60\% of the total time in the naive HNSW search process, as shown in Fig. \ref{fig: sparse cost}.
\end{enumerate}

To address the above challenges, we propose a distribution alignment method for dense and sparse vectors, and optimize the computation of sparse vectors during index construction and search. Specifically, for the first challenge, we statistically analyze the distributions of sparse and dense vector distances through pre-sampling to determine the optimal fusion weights. For the second challenge, we introduce an adaptive two-stage index construction and search strategy that evolves from dense to hybrid. This approach exploits our finding of a significant positive correlation between sparse and dense distances and little contribution of sparse vectors at the initial stage. Leveraging this, we can eliminate a large number of sparse distance calculations while maintaining accuracy. Additionally, we perform pruning on sparse vectors to further accelerate sparse computations. 

Overall, the contributions of this paper are as follows:
\begin{enumerate}
    \item We propose a method for aligning sparse and dense vector distributions that effectively combines both types of similarity, thereby enhancing the accuracy of recall results. Our experiments demonstrate that this approach can improve the recall accuracy by 1\%$\sim$9\%.
    \item We verify the potential performance of graph index for hybrid vector retrieval and further introduce an adaptive two-stage approach to reduce sparse vector computations. Additionally, we implement vector pruning to speed sparse computations. Compared to a naive implementation, our methods achieve a $\sim2.1\times$ speedup.
    \item We conducted thorough experiments on mainstream text retrieval datasets and a variety of dense and sparse embedding models. With the powerful performance of HNSW and our optimization method, we improve the end-to-end Queries Per Second (QPS) by 8.9$\times$$\sim$11.7$\times$ at equal accuracy compared with existing methods.
\end{enumerate}

\section{Preliminaries and Related Work} \label{sec: Preliminaries}

\subsection{Nearest Neighbor Search Algorithm}
For sparse vectors search, inverted index-based algorithms are commonly used, such as WAND and its variants \cite{broder2003efficient, dimopoulos2013optimizing}. Recently, researchers have proposed improved algorithms (e.g. LinScan and Sinnamon \cite{bruch2023approximate}) for model-based sparse vectors. These methods leverage vector sparsity to reduce computation and therefore are not suitable for dense vectors. For dense vectors search, a wide range of algorithms have been developed, including hash-based methods \cite{datar2004locality, shrivastava2014asymmetric}, Product-Quantization (PQ) \cite{guo2020accelerating, johnson2019billion}, Inverted File index \cite{douze2024faiss}, and Graph-based ANNS (GANNS) \cite{hajebi2011fast, malkov2018efficient}. These methods achieve efficient ANNS by calculating only a subset of candidates, or by approximating the calculation through quantization. However, the high dimensionality of sparse vectors makes it hard to build efficient index using IVF or hash-based methods. But for GANNS, because of the high connectivity of gragh, it has the potential to maintain excellent performance in high-dimensional space.

In this study, we adopt the HNSW, one of the most popular GANNS algorithms, as the base algorithm for hybrid vector search. HNSW constructs a graph with multiple layers, establishing connections between elements based on proximity. The search starts from the top layer, performing \textit{1-greedy search} until the nearest neighbor is found. This node becomes the entry point for the next layer. At the bottom layer (L0), HNSW performs a \textit{best-first beam search} for final candidates. The trade-off between search time and accuracy is controlled by the parameter "ef", which determines the length of the candidate queue.

\subsection{Hybrid Vector Search}
The prevailing and straightforward method for hybrid vector search is to perform sparse and dense recall separately, followed by a hybrid sorting \cite{bruch2023analysis, chen2022out}. But the two disconnected recall systems introduce data synchronization overhead and can impact scalability and accuracy. To address these issues, recent research has proposed the Sinnamon algorithm \cite{bruch2023bridging}, which projects high-dimensional sparse vectors into lower-dimensional denser vectors using a hashing function. The resulting vector is then concatenated with the dense vector, and indexes are constructed using the IVF algorithm for hybrid retrieval. However, the hashing process introduces additional errors. And the IVF method performs worse than GANNS in large-scale datasets. In addition, the LADR algorithm \cite{kulkarni2023lexically} leverages sparse retrieval results to generate the entry point for dense vector search on the index graph. However, this method primarily focuses on dense recall and does not achieve true hybrid retrieval.

In this paper, we suppose that dense models embed queries and candidate texts into $N$-dimensional vectors, denoted as $q^{d}, d^{d} \in \mathbb{R}^N$ respectively. And sparse models embed them into $M$-dimensional vectors, denoted as $q^{s}, d^{s} \in \mathbb{R}^M$, where $M >> N$. Then we can write the hybrid vectors as $q = q ^{d} \oplus q ^{s}, d = d ^{d} \oplus d ^{s} \in \mathbb{R}^{N+M} $, where $\oplus$ denotes concatenation. We denote the distance function between vectors as $f(x_1,x_2)$ and adopt the Inner Product (IP) distance function, \textit{i.e.} $f(x_1,x_2)=1-\left \langle x_1,x_2 \right \rangle$. 
To combine the dense distance and sparse distance, most works adopt the linear weighted method to calculate the hybrid distances \cite{gao2021complement, lee2023complementarity, wang2021bert, luan2021sparse}: 
\begin{equation} \label{equ: hybrid distance}
    f_h(q, d) = \alpha \cdot f(q^d, d^d) + (1-\alpha) \cdot f(q^s, d^s)
\end{equation}
where $f_{h}$ is the distance computation function for hybrid vectors and the hyperparameter $\alpha$, which is called dense weight, controls the relative importance of the dense and sparse distances.
Previous studies have also proposed the Reciprocal Rank Fusion (RRF) method for hybrid scores computation \cite{chen2022out}. However, as our approach does not follow the two-route retrieval method, we lack the ranking positions of candidates. Therefore, we adopt the linear weighted method, which has been shown to achieve excellent performance \cite{bruch2023analysis}.
The objective of hybrid vector search is to identify the set $\mathcal{S}$ of top-\textit{k} vectors with the minimal hybrid distance to query vector $q$ from the candidate dataset $\mathcal{D}$:
\begin{equation}
    \mathcal{S} = arg \min_{\textbf{d}\in D}^{(k)} f_h(q,d) 
\end{equation}

\section{Dense and Sparse Space Alignment} \label{sec: Alignment}
In this section, we introduce the the distribution alignment method to determine the optimal fusion weight and achieve high accuracy.

\begin{figure}[!tp] 
  \centering
  \includegraphics[width=\linewidth]{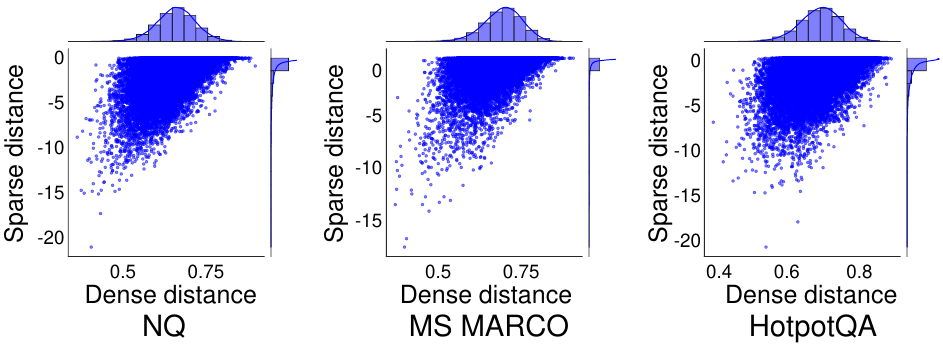}
  \caption{The distribution of dense distance and sparse distance on NQ, MS MARCO, and HotpotQA datasets embedded by BGE model and SPLADE model.}
  \vspace{-10pt}
  \label{fig: IP_distribute}
\end{figure}

Typically, dense embedding vectors are normalized to unit magnitude, with IP distances ranging from zero to one \cite{li2023general, bge_embedding}. In contrast, sparse vectors have indeterminate magnitudes, often exceeding one. Due to their high dimensionality and sparsity, sparse vectors of unrelated texts likely yield an IP of zero (IP distance of one), while those with high keyword matches may produce an IP greater than one (IP distance below zero). Note that we adopt linear weighting for sparse and dense distances, the negative sparse distance does not affect the final hybrid distance ranking. We analyze the vector distance distribution on real datasets by sampling and the results are shown in Fig. \ref{fig: IP_distribute}. The distribution space and probability of sparse and dense distance exhibit significant differences, which means that same absolute distance between sparse versus dense represents very different levels of similarity. So it is hard to choose the optimal fusion weight, as illustrated in Fig. \ref{fig: dense weight}. Unlike the two-route retrieval method, which can normalize scores after obtaining separate search results \cite{bruch2023analysis}, our method needs to predetermine the weight for index building and hybrid search. Addressing this challenge requires carefully normalizing and aligning the distributions of sparse and dense vectors.

Firstly, we normalize the sparse vector of candidate texts, \textit{i.e.}  dividing by the maximum value of magnitude: 
\begin{equation}
    d^{s}_{norm} = \frac{d^{s}}{\max_{d\in \mathcal{D}}\left \| d^{s} \right \| }
\end{equation}
This is because the maximum IP of dense vectors is 1, which reflecting the highest level of similarity. We hope to limit the maximum IP of sparse vectors to be 1 as well. The sparse query vectors are normalized by dividing by the same denominator since the maximum magnitude of the query sparse vectors cannot be predetermined.

After normalization, the further alignment is performed by pre-sampling and analyzing the IPs distribution. Since our objective is to identify the top-\textit{k} vectors with minimal distances, what is crucial is not the absolute values of distances but the relative differences. And we should focus on vectors with small distance. Therefore, we align the distance differences among vectors with small distances. Specifically, we compute the differences between the minimum IP distance and the top-1\% IP distance from the sampled data for both sparse and dense vectors. The scale factor $\gamma$ for sparse vectors is determined by the ratio of the two differences: 
\begin{equation}
    \gamma =\frac{f(q^d, d^d_{top1\%})-f(q^d, d^d_{min})}{f(q^s_{norm}, d^s_{norm, top1\%}))-f(q^s_{norm}, d^s_{norm, min})} 
\end{equation}
The distributions of sparse and dense distances exhibit a certain degree of consistency across different queries \cite{bruch2023analysis}, allowing us to analyze a small subset of queries through pre-sampling. Experimental on real datasets has shown that by randomly selecting less than 1\% of queries and documents, we can obtain a reasonably accurate distance distribution.

After alignment, the optimal fusion weights tend to be around 0.5, as shown in Fig. \ref{fig: dense weight}. To determine the optimal weight more precisely, we select values around 0.5 as the fusion weights to perform NNS on a small pre-sampled portion of the query vectors and choose the optimal weight based on search accuracy. The finally hybrid distance formula is: 
\begin{equation} \label{equ: hybrid distance optimal}
    f_h(q, d) = \alpha \cdot f(q^d, d^d) + (1-\alpha) \cdot \gamma \cdot f(q^s_{norm}, d^s_{norm})
\end{equation}
Compared to directly adopting the weight of 0.5, our method can improve recall accuracy by approximately 1\% to 9\%.

\section{Sparse computation optimization} \label{sec: sparse optimization}
Thanks to the efficient graph index in high-dimensional space, the original HNSW algorithm has achieved better performance than existing algorithms in hybrid vector retrieval. However, the sparse computing overhead causes the algorithm's efficiency to be worse than pure dense recall at low accuracy levels.
To address this, we introduce two ways to reduce the sparse computation overhead: 1) reducing the number of sparse vector calculations by adopting a dense-to-hybrid two-stage strategy, and 2) accelerating individual computations by pruning non-zero elements in sparse vectors. 

\subsection{Adaptive Two-stage Computation Strategy} \label{sec: two-stage}
\begin{figure}[!tp] 
  \centering
  \includegraphics[width=0.9\linewidth]{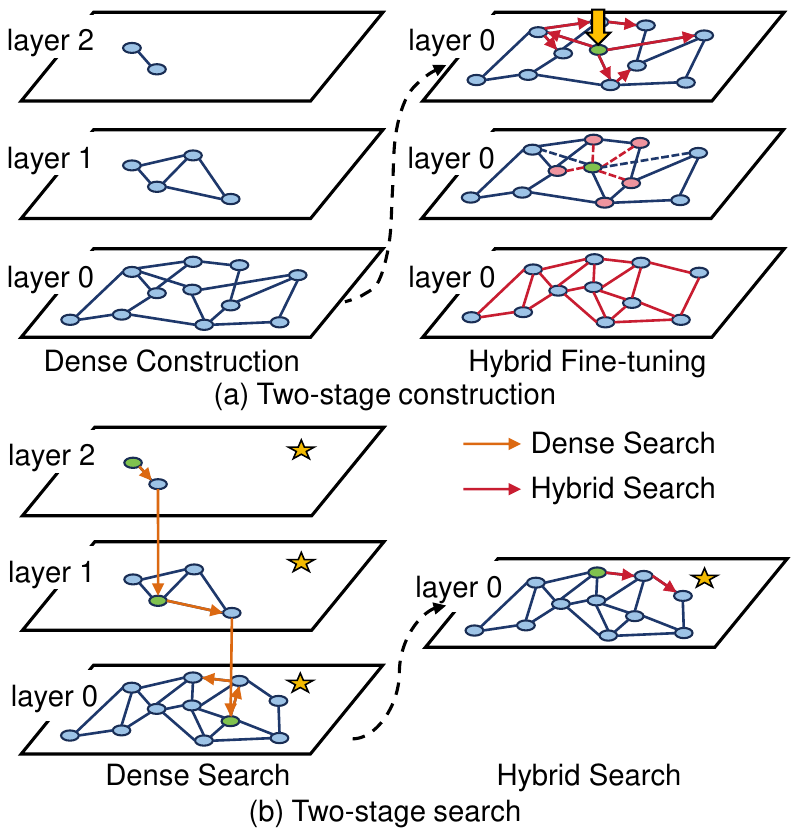}
  \caption{Illustration of two-stage graph construction and two-stage search processes.}
  \label{fig: two stage}
\end{figure}

The sparse distance captures the degree of keyword matching, while the dense distance reflects semantic similarity. Empirical evidence suggests that these two types of similarities typically exhibit a positive correlation in practical scenarios. We can observe that in Fig. \ref{fig: IP_distribute}. This observation has inspired us to approximate the hybrid distance by the dense distance alone in situations where precise hybrid distances are not necessary. 
In GANNS algorithms, the initial phases of index construction and search do not necessitate high precision in distance measurements. During the search stage, we start with an extensive coarse search before transitioning to a precise local search. Therefore, sparse distance information is not necessary in the initial stage. So we propose the dense-to-hybrid two-stage strategy for both construction and search, as illustrated in Fig. \ref{fig: two stage}. 

\begin{algorithm}[!tp]
    \renewcommand{\algorithmicrequire}{\textbf{Input:}}
	\renewcommand{\algorithmicensure}{\textbf{Output:}}
    \caption{Two-stage Construction Algorithm}
    \label{alg: construction}
    \begin{algorithmic}[1]
        \REQUIRE vector set $\mathcal{D}$, number of neighbors \textit{M}, length of dense candidate queue $\textit{cef-dense}$, length of hybrid candidate queue $\textit{cef-hybrid}$, construction dense-weight $\alpha$
        \ENSURE Multi-layer graph $\textit{hnsw-hybrid}$
        \STATE \# Stage 1: Dense Construction
        \STATE build dense graph $ \textit{hnsw-dense} \gets HNSW_{construction}(\mathcal{D}, M, $\\
        \hspace{\algorithmicindent} $\textit{cef-dense}, \textit{distance = dense-only)}$
        \STATE \# Stage 2: Hybrid Fine-tuning
        \FOR{$d \in \mathcal{D}$}
        \STATE $R \gets HNSW_{search}(d, ep=d, ef = \textit{cef-hybrid},$\\ 
        \hspace{\algorithmicindent} $weight = \alpha, \textit{distance = hybrid})$
        \STATE $\textit{neighbors}[d] \gets$ heuristic select \textit{M} neighbors from $R$ 
        \ENDFOR
        \STATE $\textit{hnsw-hybrid} \gets$ using $\textit{neighbors}$ to update $\textit{hnsw-dense} ~L0$
        \STATE \textbf{return} Multi-layer graph \textit{hnsw-hybrid}
    \end{algorithmic}
\end{algorithm}

\begin{algorithm}[!tp]
    \renewcommand{\algorithmicrequire}{\textbf{Input:}}
	\renewcommand{\algorithmicensure}{\textbf{Output:}}
    \caption{Two-stage Search Algorithm}
    \label{alg: search}
    \begin{algorithmic}[1]
        \REQUIRE query vector $q$, number of search results \textit{top-k}, length of candidate queue \textit{sef}, dense search stop threshold $\tau \textit{-dense} $, hybrid search stop threshold $\tau \textit{-hybrid} $, dense-weight $\alpha$
        \ENSURE top-k nearest vectors to $q$
        \STATE \textit{C} = candidate queue
        \STATE \textit{R} = nearest neighbor queue
        \STATE $ep_i$ = enter point on the layer $i$
        \STATE \#Stage 1: Dense Search
        \STATE \textit{C} $\gets$ insert $ep_l$
        \FOR{$i \gets l ~to~ 1$}
            \STATE \textit{R}, \textit{C} $\gets HNSW_{search}(q, ep=ep_i, ef=1,$\\
            \hspace{\algorithmicindent} $distance=dense, layer=i)$ 
            \STATE $ep_{i-1} \gets$ get the nearest vector from \textit{R}
        \ENDFOR
        \STATE \textit{R, C} $\gets HNSW_{search}(q, ep=ep_0, ef=sef, $ \\ 
        \hspace{\algorithmicindent} $distance=dense, layer=0, threshold=\tau \textit{-dense})$
        \STATE \#Stage 2: Hybrid Search 
        \STATE \textit{R, C} $\gets$ update the hybrid distance of vectors in \textit{R, C}
        \STATE $ep_0 \gets$ get the nearest vector from \textit{R}
        \STATE $R \gets HNSW_{search}(q, ep = ep_0, ef = \textit{sef}, weight = \alpha, $\\
        \hspace{\algorithmicindent} $distance=hybrid, layer=0, threshold=\tau \textit{-hybrid})$ 
        \STATE \textbf{return} \textit{top-k} nearest vectors from \textit{R}
    \end{algorithmic}
\end{algorithm}

\subsubsection{Construction Algorithm}
\label{sec: Construction Algorithm}
We extend the original HNSW construction algorithm to comprise two phases: a dense-only construction phase and a hybrid fine-tuning phase. The algorithm is presented as Algorithm \ref{alg: construction}. 

During the dense-only construction process, we follow the original HNSW construction algorithm, which using dense distances to select neighbors. The parameter \textit{cef-dense} controls the trade-off between construction time and graph quality. When sequentially inserting nodes into the graph, we store both sparse vectors and dense vectors in the graph. The fixed-length dense vectors are stored in an array format. Since the number of non-zero elements in each sparse vector is uncertain, we adopt a storage format similar to the Compressed Sparse Row (CSR) format. 
This format saves the values and corresponding index positions of non-zero elements separately in two arrays, namely values and indices. Additionally, the starting pointers of each vector are saved in a third array called index pointers. Compared to the original dense HNSW graph index, the additional storage overhead of the hybrid index is the sparse vector component.

After initially building the graph using only dense distances, we refine it based on the sparse-dense hybrid distance. The goal is to further improve the quality of the graph index. Specifically, we focus on enhancing the connections at the lowest level ($L0$) of the multi-layer graph. For each node at this level, we perform a hybrid search using the node itself as the entry point ($ep$). The aim is to find better neighbors by leveraging both sparse and dense distances. The \textit{cef-hybrid} parameter controls the tradeoff between search time and accuracy. The weight parameter $\alpha$ determines the relative importance of sparse and dense distances in the hybrid distance metric. We then update the neighbor list of each node by heuristically selecting neighbors from the hybrid search results.

The initial phase calculates only dense distances, which is significantly faster than the naive approach of computing the hybrid distance. While the two-stage method includes an additional refinement phase using hybrid distances, this second stage does not add substantial time overhead. This is because the initial neighbor graph eliminates the need for extensive searches during the refinement. By setting the \textit{cef-hybrid} parameter to a relatively small value, e.g. 16 or 32, a satisfactory graph can be obtained. As a result, the two-stage construction method achieves $\sim$2$\times$ acceleration compared to the naive implementation.

\subsubsection{Search Algorithm}
\label{sec: Search Algorithm}
The HNSW search process has two phases: the upper layer graph search and the lowest layer graph search. For the upper layers, we use the standard HNSW search and only calculate dense distances. In the lowest layer, we employ the two-stage strategy, as shown in Algorithm \ref{alg: search}.

In the initial stage of the lowest layer graph search, we use dense distance for a rough search. Once the search satisfies the termination condition regulated by the parameters $\tau \textit{-dense}$ and $sef$, the dense search is stopped. When transitioning from the dense search to the hybrid search, we update the distance values in the nearest neighbor queue $W$ and the candidate queue $C$, replacing the dense distances with hybrid distances. We then extract the node with the minimum hybrid distance from $C$ as the entry point and perform the hybrid search. The hybrid search is stopped when the condition, governed by the parameters $\tau \textit{-hybrid}$ and $sef$, is met.

Unlike conventional HNSW algorithms, our two-stage search uses a stopping condition determined by both the $sef$ parameter and the $\tau$ parameter. The $sef$ parameter controls the length of the candidate queue. In our algorithm, it affects the transition from the dense search to the hybrid search, as the starting point of the hybrid search is selected from the candidate queue. The nodes present in the candidate queue constitute the initial set for the hybrid search stage. A longer candidate queue can enhance the entry point and improve recall accuracy. However, a larger $sef$ can also lead to a looser stopping condition and longer search times. To address this, we introduce the $\tau$ parameter to further control the stopping condition. The search terminates when the number of updated points in the candidate queue during the current iteration is less than the product of the queue length ($sef$) and (1-$\tau$), where $\tau$ is between zero and one. When $\tau$ is one, the condition is consistent with the original algorithm, and the search terminates more quickly as $\tau$ decreases. We do not use the stopping condition from prior work, which employs the top-$sef$ distance multiplied by a parameter $\tau$ as a distance threshold. This is because the sparse distance may not be non-negative, making the condition potentially erroneous. The three parameters, $sef$, \textit{$\tau$-dense}, and \textit{$\tau$-hybrid}, control the tradeoff between search speed and accuracy together. With two-stage search, the number of sparse distance calculations can be reduced by 3$\times\sim$7$\times$.

\begin{figure}[!tp] 
  \centering
  \includegraphics[width=0.9\linewidth]{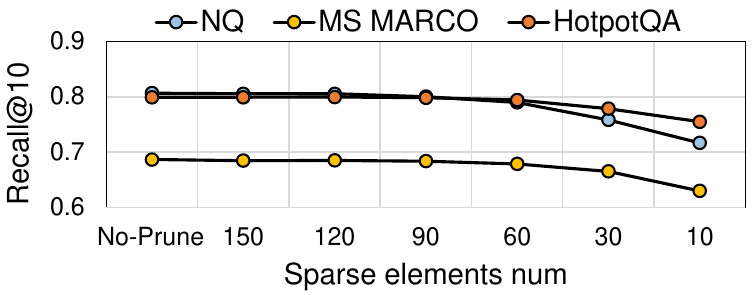}
  \caption{Effect of sparse vector pruning on recall accuracy for vectors embedded by Bge\&Splad model on NQ, MSMARCO, and HotpotQA datasets.}
  \vspace{-10pt}
  \label{fig: prune recall}
\end{figure}

\begin{table}[!tp]
\caption{Information of datasets, including the number of base vectors, the dimensions (Dim) and the average number of non-zero elements for candidate sparse vectors (in parentheses is for queries sparse vectors).}
\resizebox{\linewidth}{!}{
\label{tab:datasets}
\begin{tabular}{c|c|cc|cc|c|c}
\toprule
\multirow{2}{*}{Dataset} & \multirow{2}{*}{\makecell{Document \\ Count}} & \multicolumn{2}{c|}{SPLADE}      & \multicolumn{2}{c|}{BM25}         & \multirow{2}{*}{\makecell{BGE \\ Dim}}  & \multirow{2}{*}{\makecell{GTE \\ Dim }} \\ \cmidrule{3-6}
                         &                                 & \makecell{element \\ num } & Dim & \makecell{element \\ num } & Dim  &                           &                          \\ \midrule
NQ                       & 2.68M                           &153 (49) & \multirow{3}{*}{30522} & 40 (5)   & \multirow{3}{*}{4295M} & \multirow{3}{*}{1024}     & \multirow{3}{*}{768}     \\
MS MARCO                 & 8.8M                            & 127 (49)    &                    & 26 (5)   &                        &                           &                          \\
HotpotQA                 & 5.23M                           & 131 (59)    &                    & 24 (9)   &                        &                           &                          \\ \bottomrule
\end{tabular}
}
\end{table}

\begin{table*}[!tp]
\caption{Comparison of graph index construction time and quality across different methods on NQ, MS MARCO, and HotpotQA datasets embedded by Bge\&Splade models.}
\resizebox{\linewidth}{!}{ 
\label{tab:index construction}
\begin{tabular}{c|ccc|ccc|ccc}
\toprule
             & \multicolumn{3}{c|}{NQ}                          & \multicolumn{3}{c|}{MS MARCO}                    & \multicolumn{3}{c}{HotpotQA}                     \\
Method       & \makecell{Build Time \\ (us)} & \makecell{QPS \\ (recall@10$\sim$0.99)} & \makecell{recall@10 \\ (QPS$\sim$500)} & \makecell{Build Time \\ (us)} & \makecell{QPS \\ (recall@10$\sim$0.96)} & \makecell{recall@10 \\ (QPS$\sim$250)} & \makecell{Build Time \\ (us)} & \makecell{QPS \\ (recall@10$\sim$0.94)} & \makecell{recall@10 \\ (QPS$\sim$167)} \\ \midrule
Naive Hybrid &  3.11E+09 (2.10$\times$)   &   117.33 &   0.933 &  9.51E+09 (2.19$\times$)  &    167.61 &    0.946 &  6.82E+09 (2.21$\times$)  &    83.33 &   0.89  \\
Dense        &  1.11E+09 (0.75$\times$)   &   85.35  &   0.925 &  3.52E+09 (0.81$\times$)  &    128.04 &    0.933 &  2.65E+09 (0.82$\times$)  &    68.43 &   0.87  \\
Opt Hybrid   &  1.48E+09 (1.00$\times$)   &   114.97 &   0.932 &  4.35E+09 (1.00$\times$)  &    162.34 &    0.943 &  3.24E+09 (1.00$\times$)  &    80.6  &   0.89  \\ \bottomrule
\end{tabular}
}
\end{table*}

\subsection{Sparse Vector Pruning}
Graph index construction and search require discontinuous access to candidate vectors. Therefore, we use an indices-value pair format to store sparse vectors, instead of an inverted index. The computation time of the IP distance between sparse vectors is positively correlated with the number of non-zero elements. This computation time significantly exceeds that of dense vectors with the same number of non-zero elements, as shown in Fig. \ref{fig: sparse cost}. Importantly, the IP computation has the property that elements with larger values are more important, while those with smaller values are less important. We leverage this property to design a pruning strategy, where we trim the elements with smaller values. According to experimental results (Fig. ~\ref{fig: prune recall}), more than 40\% of the data in the sparse vector can be pruned without significantly reducing the recall. As a result, the IP computation of sparse vectors can be accelerated by $\sim 1.4 \times$.

\section{Experiments}
\subsection{Experiment Setup}
\subsubsection{Hardware Configuration}
We conducted all the experiments on a CPU server, which is equipped with an Intel(R) Xeon(R) Platinum 8260 CPU(24 cores, running at 2.40 GHz). The operating system is Debian GNU/Linux 9.13. To save time, we builded indexes with 10 threads. And during the search process, we all employed one thread to compare the algorithms. 

\begin{figure}[!tp] 
  \centering
  \includegraphics[width=0.9\linewidth]{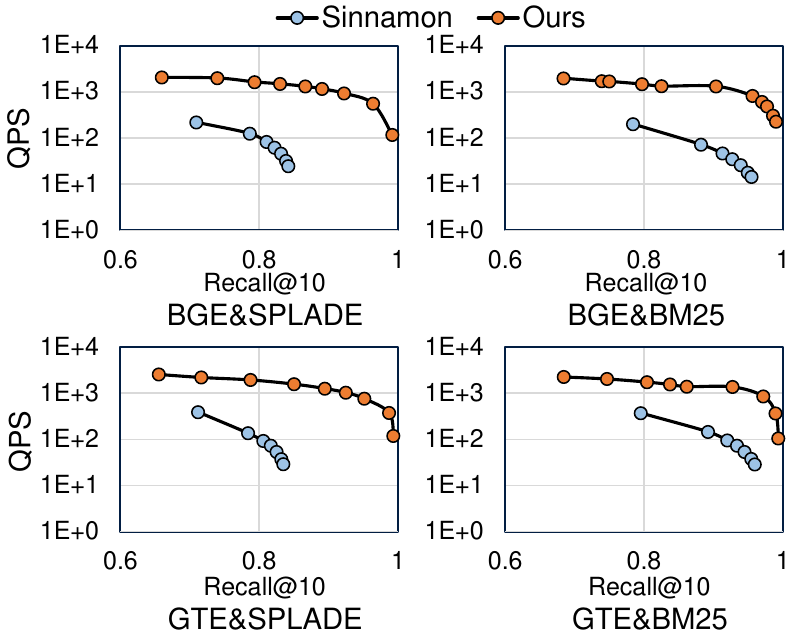}
  \caption{ QPS and Recall@10 against brute force on NQ dataset.}
  \label{fig: recall_QPS_gt}
\end{figure}

\subsubsection{Datasets}
The datasets and embedding models that we used are shown in the Table.~\ref{tab:datasets}. For dense embedding models, we chose BGE-large\footnote{https://huggingface.co/BAAI/bge-large-en-v1.5} \cite{bge_embedding} and GTE-base\footnote{https://huggingface.co/thenlper/gte-base} \cite{li2023general}, which are among the SOTA dense models for text retrieval \footnote{https://huggingface.co/spaces/mteb/leaderboard}. For sparse methods, we chose the SPLADE model\footnote{https://github.com/naver/splade} \cite{formal2021splade} and the BM25 method\footnote{https://github.com/pinecone-io/pinecone-text}. The former is a sparse embedding model, while the latter is a classic term frequency-based method. The average number of non-zero elements and the dimension of sparse vectors are different for different models and datasets. We present the characteristics of vectors derived from various datasets using different models in Table~\ref{tab:datasets}.

\subsubsection{Baselines}
To evaluate our index construction algorithm, we compared it to the original HNSW methods, including the naive hybrid algorithm and the dense algorithm. The naive hybrid algorithm uses the hybrid distance metric to construct the index. The dense algorithm only uses the dense distance, but stores both vectors.

\begin{figure*}[!tp] 
  \centering
  \includegraphics[width=0.95\linewidth]{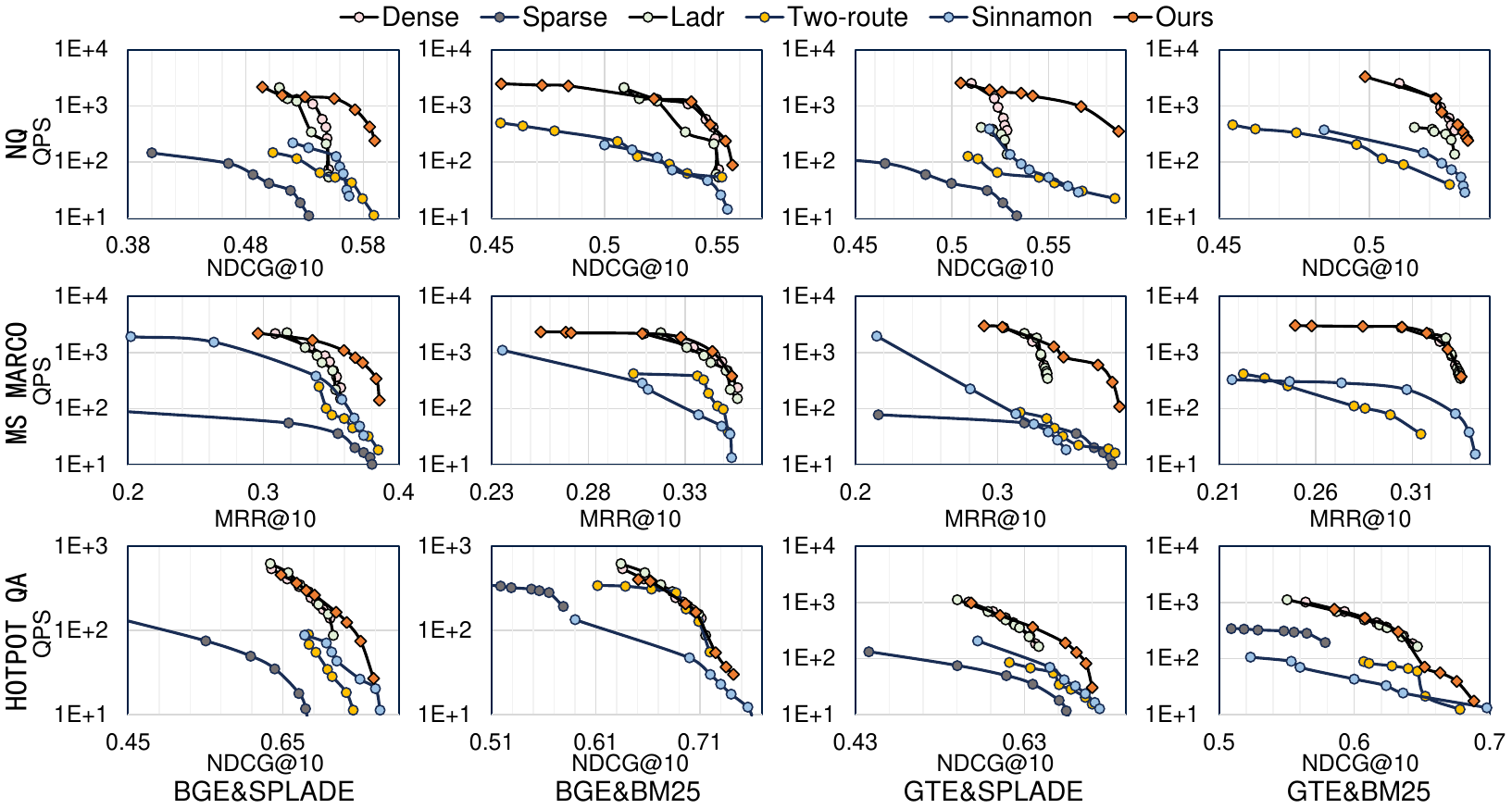}
  \caption{QPS and end-to-end Recall@10 on NQ, MS MARCO and HotpotQA datasets.}
  \label{fig: recall_QPS}
\end{figure*}

\begin{figure}[!tp] 
  \centering
  \includegraphics[width=0.95\linewidth]{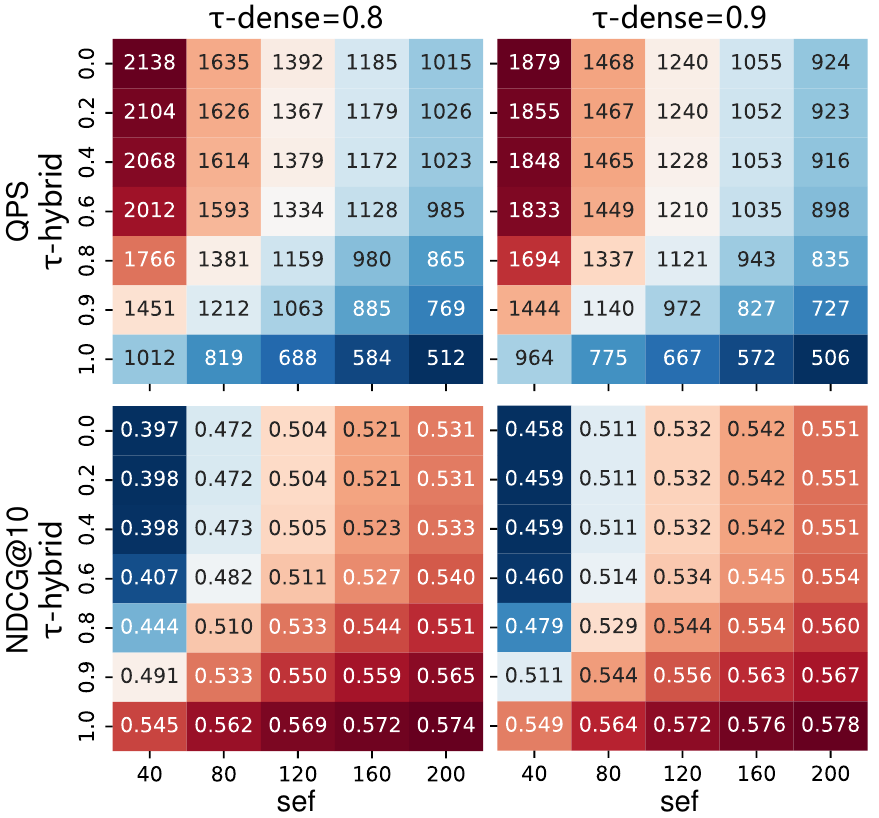}
  \caption{Comparison of our algorithm performance on NQ embedded by Bge\&Splade models while varying $\tau\mbox{-}dense$, $\tau\mbox{-}hybrid$, and $sef$.}
  \label{fig: parameter}
\end{figure}

For the search stage, we compared to two existing hybrid ANNS approaches: the two-route retrieval strategy and the Sinnamon algorithm \cite{bruch2023bridging}. In our experiments, the two-route strategy employs HNSW for dense vector retrieval, while using WAND \cite{broder2003efficient} for searching BM25 sparse vectors and LinScan \cite{bruch2023approximate} for searching SPLADE sparse vectors. The Sinnamon algorithm projects sparse vectors into a lower-dimensional dense space using hashing and then uses an IVF-based algorithm to search the dense space. We reproduced this method according to the algorithm \cite{bruch2023bridging}, while setting the hash mapping dimension to 1024 and the number of clusters to 1000. We also compared to pure dense vector retrieval, pure sparse vector retrieval, and the LADR algorithm \cite{kulkarni2023lexically}, which uses sparse BM25 vectors to quickly find starting points for dense searches.

\subsubsection{Parameters setting}
For index construction, we set the parameter $m$ to 32 and $cef$ to 200 by default. And we set $cef\mbox{-}hybrid$ to 32 in our two-stage construction algorithm. In the search stage, we vary the key parameters: the $sef$ ranges from 10 to 200, $\tau\mbox{-}dense$ is set between 0.6 and 1.0, and $\tau\mbox{-}hybrid$ is set between 0.0 and 1.0. 

\subsection{Experiment Results}

\subsubsection{Graph Index Construction}
The comparison of the construction time and quality of the graph index is presented in Table \ref{tab:index construction}. We executed hybrid searches on these indices and compute recall@10 against brute force results as accuracy metric. The results indicate that our approach, denoted as Opt Hybrid, enhances the construction efficiency by approximately 2.1$\times$ compared to the naive hybrid method and delivers superior search performance when compared to the dense method.  

\subsubsection{Recall against Brute Force}
We tested the QPS and accuracy of retrieval results against brute-force search between our methods and Sinnamon. The experiment is conducted on NQ dataset, using the BGE or GTE model to generate dense vectors, and using the SPLADE model or BM25 algorithm to generate sparse vectors. We select recall@10 as the accuracy metric.

Fig.\ref{fig: recall_QPS_gt} shows the performance comparison results. Due to the errors introduced by hash mapping and the inherently poor performance of IVF, Sinnamon performed relatively poorly. Our method achieved over 10 $\times$ higher QPS compared to the Sinnamon algorithm, while can achieve higher recall accuracy.

\subsubsection{End-to-end Performance}
To further validate the performance of hybrid vector search in information retrieval, we conducted end-to-end experiment. We calculated the retrieval performance based on the test data provided in real retrieval datasets. We chose mrr@10 as metric for MS MARCO and ndcg@10 for NQ and HotpotQA. The experimental results are shown in Fig. \ref{fig: recall_QPS} and Table \ref{tab:recall_QPS}.

Overall, our algorithm provides the best QPS-vs-accuracy performance. The two-route retrieval method has a long latency and low QPS because of the need for separate sparse and dense retrieval. Compared with the Sinnamon algorithm and two-route method, our method can achieve 8.9$\times$$\sim$11.7$\times$ QPS at equal retrieval accuracy. Althrough dense retrieval and LADR algorithm can achieve similar performance at lower accuracy thresholds, they are unable to attain the same high accuracy levels as hybrid search.

\begin{table*}[!tp]
\caption{Comparison of QPS across different methods at two accuracy points on NQ, MS MARCO, and HotpotQA datasets. Datasets are embedded using Bge/Gte models for dense and Splade/BM25 models for sparse. Results that cannot be matched to the accuracy point are indicated with "-".}
\resizebox{0.9\linewidth}{!}{
\label{tab:recall_QPS}
\begin{tabular}{cc|cccc|cccc|cccc}
\toprule
&             & \multicolumn{4}{c|}{NQ (ndcg@10)}                           & \multicolumn{4}{c|}{MS MARCO (mrr@10)}                     & \multicolumn{4}{c}{HotpotQA (ndcg@10)}                     \\
&Method       & \makecell{BGE\&\\SPLADE} & \makecell{BGE\&\\BM25} & \makecell{GTE\&\\SPLADE} & \makecell{GTE\&\\BM25} & \makecell{BGE\&\\SPLADE} & \makecell{BGE\&\\BM25} & \makecell{GTE\&\\SPLADE} & \makecell{GTE\&\\BM25} & \makecell{BGE\&\\SPLADE} & \makecell{BGE\&\\BM25} & \makecell{GTE\&\\SPLADE} & \makecell{GTE\&\\BM25} \\
\midrule \midrule
&Accuracy     & 0.55 & 0.53 & 0.53 & 0.52 & 0.35 & 0.35 & 0.33 & 0.32 & 0.70 & 0.71 & 0.64 & 0.64 \\
\midrule
\multirow{6}{*}{\rotatebox{90}{QPS}} 
&Dense     & 73.15   & 1078.45 & 336.72  & 1350.12 & 687.59  & 687.59 & 874.16  & 1851.92 & 177.54 & 142.26 & 225.18 & 225.18 \\
&Sparse    & -       & -       & 15.942  & -       & 42.32   & -      & 48.31   & -       & -      & -      & 35.02  & -      \\ 
&Ladr      & 53.55   & 871.43  & 137.68  & 382.75  & 478.34  & 478.34 & 943.57  & 1861.25 & 184.69 & 148.55 & 233.44 & 233.44 \\ 
&Two-route & 57.31   & 87.66   & 58.12   & 59.23   & 77.55   & 98.43  & 68.43   & 21.56   & 44.54  & 128.43 & 61.35  & 66.70  \\ 
&Sinnamon  & 136.08  & 72.04   & 135.34  & 145.25  & 220.78  & 39.65  & 44.05   & 145.33  & 76.31  & 42.39  & 96.47  & 22.39  \\ 
&Ours      & 1378.38 & 1178.82 & 1686.18 & 1369.24 & 1365.91 & 686.87 & 1672.77 & 1852.74 & 216.90 & 135.45 & 360.38 & 231.68 \\ 
\midrule \midrule
&Accuracy     & 0.58 & 0.55 & 0.57 & 0.53 & 0.38 & 0.36 & 0.38 & 0.33 & 0.74 & 0.74 & 0.71 & 0.68 \\
\midrule
\multirow{6}{*}{\rotatebox{90}{QPS}} 
&Dense     & -       & 73.15   & -       & -       & -       & 162.01 & -       & 874.16  & -      & -      & -      & -      \\
&Sparse    & -       & -       & -       & -       & 10.28   & -      & 10.28   & -       & -      & -      & -      & -      \\ 
&Ladr      & -       & 53.55   & -       & -       & -       & 109.58 & -       & 943.57  & -      & -      & -      & -      \\ 
&Two-route & 22.24   & 56.43   & 28.34   & 21.54   & 26.55   & 12.47  & 17.98   & 8.94    & 12.36  & 11.49  & 15.44  & 9.04   \\ 
&Sinnamon  & -       & 29.88   & 14.96   & 53.63   & 9.95    & -      & -       & 103.58  & 32.57  & 17.63  & 18.69  & 18.27  \\ 
&Ours      & 664.66  & 345.79  & 846.63  & 463.26  & 476.58  & 164.69 & 299.96  & 986.48  & 113.59 & 32.25  & 30.37  & 26.85  \\ 
\bottomrule
\end{tabular}
}
\end{table*}

\begin{figure}[!tp] 
  \centering
  \includegraphics[width=0.9\linewidth]{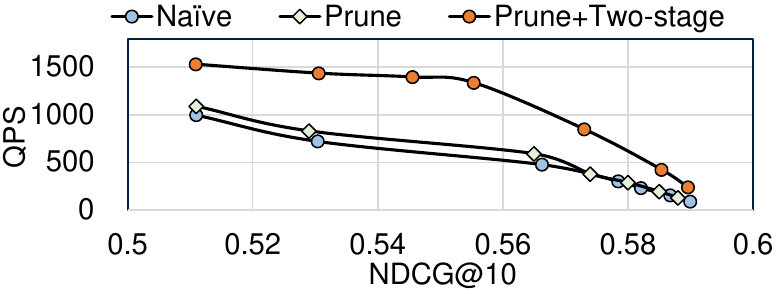}
  \caption{Ablation experiment results (QPS-vs-accuracy) on NQ dataset embedded by Bge\&Splade.}
  \label{fig: ablation}
\end{figure}

\begin{figure}[!tp] 
  \centering
  \includegraphics[width=0.9\linewidth]{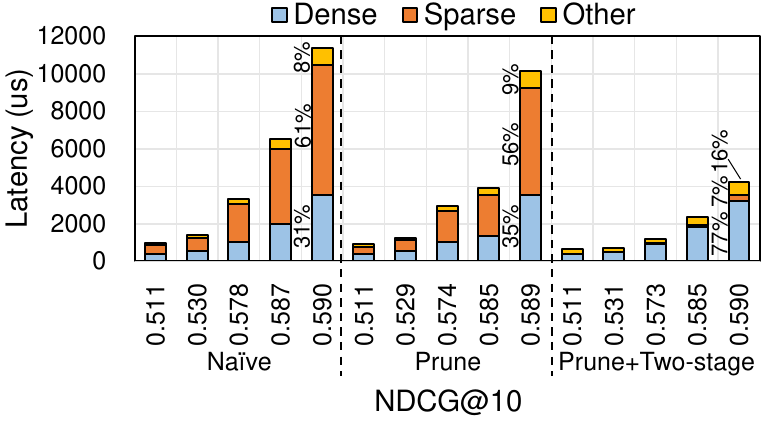}
  \caption{Ablation experiment results (latency distribution) on NQ dataset embedded by Bge\&Splade.}
  \label{fig: time distribution}
\end{figure}

\subsubsection{Effect of Parameters}
We performed an analysis of the key parameters in our proposed algorithm: the length of the candidate queue during search ($sef$) and the stop condition for dense search ($\tau\mbox{-}dense$) and for hybrid search ($\tau\mbox{-}hybrid$). Fig. \ref{fig: parameter} reports the retrieval accuracy (NDCG@10) and QPS on NQ dataset embedded by Bge\&Splade models.
We observed that as the $sef$, $\tau\mbox{-}dense$, and $\tau\mbox{-}hybrid$ increase, the results accuracy also improves, but at the cost of decreased QPS. Furthermore, when $sef$ and $\tau\mbox{-}dense$ are both large, the increase of $\tau\mbox{-}hybrid$ improves the accuracy slowly, but significantly reduces the QPS. These findings suggest that when improving result accuracy, the focus should be on increasing $sef$ and $\tau\mbox{-}dense$ first, followed by adjustments to $\tau\mbox{-}hybrid$.

\subsubsection{Ablation Studies}
We conducted an ablation experiment on NQ dataset using Bge\&Splade models to demonstrate the effect of the sparse vector pruning and two-stage search method on performance. We implemented a naive hybrid search based on HNSW, which directly replaces the dense distance calculation with the hybrid distance. The QPS-vs-accuracy performance comparison and latency analysis are shown in Fig. \ref{fig: ablation} and Fig. \ref{fig: time distribution}. 
Since the excellent performance of the HNSW index, the naive implementation already outperformed the two-stage algorithm and Sinnamon algorithm. However, due to the overheads introduced by sparse vectors, the naive implementation underperformed the pure dense retrieval at lower accuracy thresholds. By implementing sparse vector pruning, we achieved $\sim$1.2$\times$ acceleration. Combining this with the two-stage search strategy resulted in a total 2.1$\times$ speedup and outperforming pure dense retrieval in general. As shown in Fig. \ref{fig: time distribution}, the primary time bottleneck for the naive implementation is the sparse vector computation component, for which our approach is optimized.

\section{Conclusion}
Sparse-dense hybrid vector search significantly improves the accuracy of text retrieval, but efficient and  effective implementation remains challenging. The vast differences of distribution between sparse and dense vectors make it difficult to properly fuse the two distance metrics, impacting precision. We addressed this by aligning the sparse and dense vector distributions through sampling analysis, achieving 1$\sim$9\% accuracy gains. For efficient search, we validated the superior performance of the HNSW algorithm for hybrid retrieval. However, the high computational overhead of sparse vectors degraded efficiency, underperforming dense methods at low accuracy thresholds. To optimize sparse computation overhead, we propose a two-stage dense-to-hybrid strategy, taking into account the limited contribution of sparse embedding information in the initial stages. And we introduce sparse vector pruning to accelerate individual computation. Ultimately we achieve a 2.1$\times$ speedup over the naive approach. Leveraging the efficient HNSW algorithm and our optimization techniques, our hybrid retrieval method attains 8.9$\sim$11.7$\times$ throughput improvements over existing hybrid approaches at comparable accuracy levels.

\bibliographystyle{ACM-Reference-Format}
\bibliography{ref}


\end{document}